# Aggregate Model of District Heating Network for Integrated Energy Dispatch: A Physically Informed Data-Driven Approach

Shuai Lu, *Member*, *IEEE*, Zihang Gao, Yong Sun, Suhan Zhang, *Member*, *IEEE*, Baoju Li, Chengliang Hao, Yijun Xu, *Senior Member*, *IEEE*, Wei Gu, *Senior Member*, *IEEE*

***Abstract*—The district heating network (DHN) is essential in enhancing the operational flexibility of integrated energy systems (IES). Yet, it is hard to obtain an accurate and concise DHN model for the operation owing to complicated network features and imperfect measurements. Considering this, this paper proposes a physically informed data-driven aggregate model (AGM) for the DHN, providing a concise description of the source-load relationship of DHN without exposing network details. First, we derive the analytical relationship between the state variables of the source and load nodes of the DHN, offering a physical fundament for the AGM. Second, we propose a physics-informed estimator for the AGM that is robust to low-quality measurements, in which the physical constraints associated with the parameter normalization and sparsity are embedded to improve the accuracy and robustness. Finally, we propose a physics-enhanced algorithm to solve the nonlinear estimator with non-closed constraints efficiently. Simulation results verify the effectiveness of the proposed method.**

***Index Terms*—Aggregate model, district heating network, integrated energy systems, physics-informed data-driven method.**

## NOMENCLATURE

### A. Abbreviations

| | |
|---|---|
| DHN | District heating network |
| AGM | Aggregate model |
| IES | Integrated energy systems |
| STM | Supply temperature mapping |
| RTM | Return temperature mapping |
| LSE | Least squares estimator |
| HME | Huber M-estimator |

### B. Sets

| | |
|---|---|
| $\Phi_n$, $\Phi_p$ | Index set of nodes/pipelines in DHN |
| $\Phi_{ln}$, $\Phi_{sn}$ | Index set of load/source nodes in DHN |
| $\Phi_{p+}^k$, $\Phi_{p-}^k$ | Index set of pipelines flowing into/flowing out of node $k$ |
| $\Phi_p^{k,v}$ | Index set of pipelines on the path between node $k$ and node $v$ |
| $\Delta^{k,v}$ | Set of delay parameters from node $k$ to node $v$ |

### C. Parameters and Variables

| | |
|---|---|
| $\Delta t$ | Time length of interval (s) |
| $A_p^j$ | Cross-section area of pipeline $j$ (m$^2$) |
| $L_p^j$ | Length of pipeline $j$ (m) |
| $\lambda_p^j$ | Heat loss coefficient of pipeline $j$ (kW/(m·°C)) |
| $\rho_w$ | Heat medium density in DHN (kg/m$^3$) |
| $\gamma_p^j$, $R_p^j$, $\alpha_p^j$,$\eta_p^j$ | Parameters of the node method model |
| $c_w$ | Specific heat capacity of water (kJ/(kg·°C)) |
| $m_p^j$ | Mass flow rate of pipeline $j$ (kg/s) |
| $m_{src}^k$,$m_l^k$ | Mass flow rate of heat source/load at node $k$ of DHN (kg/s) |
| $m_{src}^{k,v}$ | Equivalent mass flow rate of the supply water from source node $k$ to load node $v$ |
| $m_l^{k,v}$ | Equivalent mass flow rate of the return water from load node $v$ to source node $k$ |
| $\tau_{amb}$ | Ambient temperature of DHN (°C) |
| $\tau_{s,src}^{k,t}$,$\tau_{r,src}^{k,t}$ | Supply/return temperature of heat source at node $k$ at period $t$ (°C) |
| $\tau_{s,l}^{k,t}$ ,$\tau_{r,l}^{k,t}$ | Supply/return temperature of heat load at node $k$ (°C) |
| $\tau_{s,n}^{k,t}$,$\tau_{r,n}^{k,t}$ | Supply/return temperature at node $k$ (°C) |
| $\tau_{s,in}^{j,t}$,$\tau_{s,out}^{j,t}$ | Inlet/outlet temperature of supply pipeline $j$ (°C) |
| $\tau_{r,in}^{j,t}$,$\tau_{r,out}^{j,t}$ | Inflow/outflow temperature of return pipeline $j$ (°C) |
| $\boldsymbol{\tau}_{s,src}^t$ | Supply temperature matrix of source |
| $\boldsymbol{\tau}_{r,src}^t$ | Return temperature vector of source |
| $\boldsymbol{\tau}_{s,l}^t$ | Supply temperature vector of load |
| $\boldsymbol{\tau}_{r,l}^t$ | Return temperature matrix of load |
| $\Gamma$ | The number of rows of the parameter matrix |
| $N_s$, $N_l$ | Number of source/load nodes in DHN |
| $N_p^{k,v}$ | Number of pipelines on the shortest path between node $k$ and node $v$ |
| $a_s^{k,v,i}$ ,$b_s^{k,v}$ | Aggregate parameters between node $k$ and node $v$ in supply network of AGM |
| $a_r^{k,v,i}$ ,$b_r^{k,v}$ | Aggregate parameters between node $k$ and node $v$ in return network of AGM |
| $\gamma_{agg}^{k,v}$ | Transmission delay from node $k$ to node $v$ |
| $\xi_s^{k,v}$ | Proportion of water mass from source node $k$ to that flows into load node $v$ |
| $\xi_r^{k,v}$ | Proportion of water mass from load node $v$ to that flows into source node $k$ |

This work was supported by the National Key R&D Plan Project of China under Grant 2022YFB2404000. (*Corresponding author: Wei Gu.*)

S. Lu, S. Zhang, Y. Xu, and W. Gu are with the School of Electrical Engineering, Southeast University, Nanjing 210096, China (e-mail: shuai.lu.seu@outlook.com; zhangsh_seu@163.com; yijunxu@seu.edu.cn; wgu@seu.edu.cn).

Z. Gao is with the School of Software, Southeast University, Suzhou 215123, China (e-mail: g18551638378@163.com).

Y. Sun, B. Li, and C. Hao are with the State Grid Jilin Electric Power Company, Changchun 130021, China (e-mail: sunyong_hit@163.com; libaoju1986@163.com; haochl@jl.sgcc.com.cn).

| | |
|---|---|
| $\tilde{a}^{k,v,i}_{s/r}$, $\tilde{b}^{v}_{s/r}$ | Model parameters of the AGM |
| $\tilde{\boldsymbol{a}}^{k}_{s}$, $\tilde{\boldsymbol{a}}^{v}_{r}$ | Parameter matrix of the AGM |
| $M$, $M_{trc}$ | Original/truncated horizon of regression model |
| $\boldsymbol{r}^{v}_{s}$, $\boldsymbol{r}^{k}_{r}$ | Residual vector of node $k/v$ in supply/return network |
| $r^{v,t}_{s}$, $r^{k,t}_{r}$ | Residual at time $t$ of node $k/v$ in supply/return network |
| $\hat{\sigma}^{v}_{s}$ | The scale estimate of residuals of the supply temperature at load node $v$ |
| $\kappa$ | Tuning constant of Huber M-estimator |

## I. Introduction

Today's increasing energy consumption and environmental degradation have generated a huge demand for improving energy efficiency and reducing carbon emissions. Integrated energy systems (IES), which combine various energy carriers and networks, have received much attention [1]. The district heating network (DHN) plays a vital role in IES because it can provide considerable flexibility for the system operation [2, 3]. Specifically, as an energy carrier, the heat medium in the DHN has excellent energy storage capacity, i.e., thermal inertia [4, 5]. In practice, the DHN and power system are usually managed by different operators, making their coordination difficult. On the one hand, an accurate DHN model that can describe the network characteristics is necessary to exploit the thermal inertia while ensuring heating quality. On the other hand, a cumbersome DHN model not only has the risk of exposing sensitive information about district heating systems to power system operators but also brings in excess computational burden for the operation. Therefore, it has been recognized that an accurate and concise DHN model is essential for the operation of IES.

Typically, the modeling of the DHN can be divided into two categories, i.e., the physics-based methods and the data-driven methods. The former directly employs physical laws to construct the DHN models [6, 7]. Typical examples include the element method [8], the characteristic method [9], and the node method [10]. Both the element method and the characteristic method divide the pipeline into many discrete "units" or "nodes", which need to be calculated in each calculation step, requiring a large amount of calculation. Compared to the above two models, the node method model has a faster computing speed [11]. Within it, the development of a pipeline model has two steps: first, the transmission delay is tracked by calculating the time of heat medium flowing through the pipeline; then, the heat loss in the transmission process is considered [12]. The accuracy of the node method has been verified in multiple real-world DHNs [13, 14]. The same philosophy among these methods emphasizes the detailed modeling of each pipeline. Such a model is usually unsuitable for the optimization problem of IES, such as optimal planning and economic dispatch. More specifically, these models introduce many equations and variables for pipelines, resulting in a large-scale optimization problem that is hard to solve. Besides, the power system and the district heating system are often owned by different entities, meaning that the above models will expose the detailed information of the DHN to others and thus cause privacy issues.

Facing the above challenges, more research has paid attention to the equivalent modeling techniques, which are much simpler yet effective. Some early work tried simplifying the DHN model to reduce computational burden by aggregating the pipelines. Larsen *et al.* [15, 16] first proposed a DHN aggregate method known as "the Danish method". It reduces the DHN to an equivalent simpler network of pipe segments based on some over-bold assumptions, such as constant overall heat load. Bøhm *et al.* [17] further tested the model performance under different aggregate depths based on "the Danish method" and "the German method" [18]. However, these methods have several shortcomings. First, the assumptions (e.g., constant overall heat load) in these methods usually cannot always hold, causing significant errors in the highly aggregated model [17]. Second, "the Danish method" requires detailed parameters, such as pipeline length and diameter. This inevitably places exceptionally high demands on the integrity and accuracy of the internal parameters. In summary, the challenges in the aggregate modeling of the DHN include: 1) How to derive the aggregate model of a multi-source heating network, especially the dynamic aggregate model. 2) How to calculate the parameters of the aggregate model in the absence of detailed pipeline parameters.

Unlike the physics-based methods that have been extensively explored, the related works for the data-driven ones are just a few, in which the time series method and neural networks are popular ones [19]. La Bella *et al.* [20] proposed a piece-wise Auto Regressive eXogenous back-box method for DHN modeling. This method can learn from the artificially generated data to build a model approximating the original nonlinear model. The data-driven DHN model suffers from several problems. First, although the models approximated from real-world measurements are more convincing, the measurements in practice are often scarce and incomplete. This is especially true for the inner pipelines and nodes of the DHN, which are usually not equipped with measurement units at all. Here, in most situations, only the operational data of source and load nodes are available. Second, the accuracy of data-driven methods depends on the quantity and quality of the data [19, 21].

To address them, a few recent works have noticed the above problems and attempted to develop the equivalent DHN model from a network perspective to avoid using the inner state data of DHN. Zheng et al. [22] first derived an equivalent matrix model for DHN that directly reveals the relationship between the boundary control and inner state variables. However, this model still needs the complete measurements to estimate the parameters since the inner states are kept. Zhang *et al.* [23] further derived the source-load function of DHN as the linear combination of the initial and boundary conditions based on the partial differential model of pipelines. Albeit simple, this method relies on detailed parameters to calculate the combination coefficients while the parameters estimation problem is ignored.

In summary, it is still an unresolved issue to develop a practical and interpretable DHN model in a data-driven manner for the optimization problem of IES that features low computational complexity and non-exposure of the inner states of network. The potential challenges are multifold. First, it needs to be further explored at the physics level regarding the accuracy of the DHN model without the internal network states. Second, although the DHN with some specific control strategies (e.g., the constant-flow strategy) is typically a linear system, its inverse problem, i.e., the parameter estimation, will be much more complicated because of the incomplete measurements. Also, the potential problem of low-quality measurements in engineering will bias the estimation results.

To address the abovementioned problems, we propose a physically informed data-driven aggregate model (AGM) for DHN by combining the advantages of physics and operational data. The proposed AGM has a concise mathematical form and clear physical connotation, yielding a cost-effective DHN model for the optimization problem of IES. In particular, the physical characteristics of the DHN are well exploited in the estimation of the AGM to improve the model accuracy and reduce the computational burden. Case studies based on different scales of the DHN verify the effectiveness of the proposed method. We further demonstrate the practicality of the AGM through the case studies in the economic dispatch of IES.

The main contributions are summarized as follows:

(1) Based on the constant flow control strategy and constant ambient temperature condition, we derive the AGM of the DHN that directly reveals the input-output relationship of network. The AGM consists of the supply temperature mapping (STM) of load nodes and the return temperature mapping (RTM) of source nodes, simplifying the DHN into a source-load mapping network by eliminating inner nodes.

(2) We propose a physics-enhanced method for the parameter estimation of the AGM. The accuracy and robustness of the estimator are improved by integrating the parameter constraints derived from the physical properties of the DHN, including the parameter normalization and sparsity.

(3) The proposed estimator is a non-closed, nonlinear optimization model due to the presence of a delay parameter in the variable index, making off-the-shelf solvers ineffective. To solve it, we propose the delay parameter enumeration-based iterative reweighted least squares (IRLS) algorithm, in which a successive estimation strategy of STM and RTM are introduced to avoid the combinatorial explosion problem.

The remainder of this paper is organized as follows. Section II introduces the AGM; Section III proposes the physics-informed robust parameter estimator for the AGM and the corresponding physics-enhanced solution algorithm; Section IV verifies the effectiveness of the proposed method by numerical tests; and Section V concludes this paper.

## II. Aggregate Model of the DHN

In this section, we will first briefly introduce the physical model of the DHN. Second, we will give the concept of the AGM. The detailed derivation of the AGM is then presented.

### A. Physical Model of the DHN

This paper focuses on the DHN that operates under the widely-used constant flow and variable temperature control strategy [24, 25]. From the perspective of network dynamic characteristics, the dynamic process in DHN includes fluid dynamics and thermal dynamics. Due to the significant difference in time constants between fluid and thermal dynamics, the fluid dynamic process is often ignored when calculating thermal dynamics [11, 26]. Therefore, the fluid in the network is assumed to be static in this work. The basic elements of DHN are illustrated in Fig. 1 (a). The DHN model consists of the heat transmission equation of pipelines, the energy balance equations of nodes, and the temperature fusion equations of nodes [25]. We use the node method [10] to model the heat transmission in the pipeline, as illustrated in Fig. 1 (b). Based on the node method, the outlet temperature of the pipeline considering the transmission delay and heat loss can be calculated as

$$\tau^{j,t}_{x,out} = \left(1-\eta^j_p\right)\left(\left(1-\alpha^j_p\right)\tau^{j,t-\gamma^j_p-1}_{x,in} + \alpha^j_p\tau^{j,t-\gamma^j_p}_{x,in}\right) + \eta^j_p\tau_{amb}. \quad \forall j\in\Phi_p, x\in\{s,r\} \tag{1a}$$

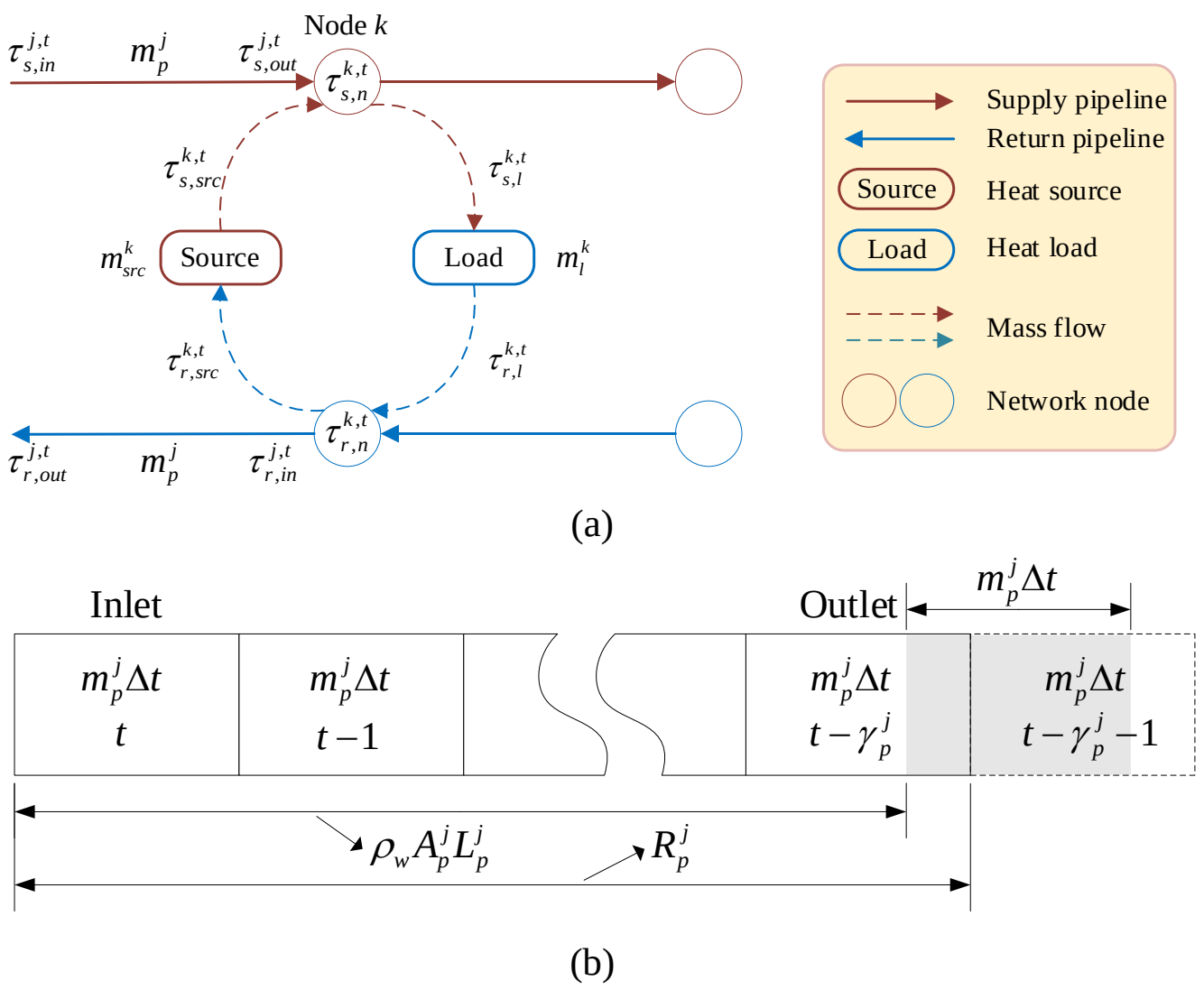

Fig. 1 The basic elements of the DHN: (a) The structure of the DHN; (b) The pipeline model based on the node method.

The parameters $\alpha^j_p$, $\eta^j_p$, and $\gamma^j_p$ can be calculated as

$$\begin{cases} \gamma^j_p = \lceil \rho_w A^j_p L^j_p / (m^j_p \Delta t) \rceil - 1 \\ R^j_p = (\gamma^j_p + 1) m^j_p \Delta t \\ \alpha^j_p = (R^j_p - \rho_w A^j_p L^j_p) / (m^j_p \Delta t) \\ \eta^j_p = 1 - \exp\left( -\dfrac{\lambda^j_p \Delta t}{\rho_w A^j_p c_w} \left(\gamma^j_p + 3/2 - \alpha^j_p\right)\right) \end{cases}, \tag{1b}$$

wherein $\lceil\cdot\rceil$ is the ceiling function.

The energy balance equations of nodes are as

$$\begin{cases} m^k_{src}\tau^{k,t}_{s,src} + \sum_{j\in\Phi^k_{p+}} m^j_p \tau^{j,t}_{s,out} = (m^k_{src} + \sum_{j\in\Phi^k_{p+}} m^j_p)\tau^{k,t}_{s,n} \\ m^k_l \tau^{k,t}_{r,l} + \sum_{j\in\Phi^k_{p-}} m^j_p \tau^{j,t}_{r,out} = (m^k_l + \sum_{j\in\Phi^k_{p-}} m^j_p)\tau^{k,t}_{r,n} \end{cases} \forall k\in\Phi_n. \tag{1c}$$

The temperature fusion equations of nodes are as

$$\begin{cases} \tau^{j,t}_{s,in}\big|_{\forall j\in\Phi^k_{p-}} = \tau^{k,t}_{s,n}, \ \tau^{j,t}_{r,in}\big|_{\forall j\in\Phi^k_{p+}} = \tau^{k,t}_{r,n} \\ \tau^{k,t}_{r,src} = \tau^{k,t}_{r,n}, \ \tau^{k,t}_{s,l} = \tau^{k,t}_{s,n} \end{cases} \forall k\in\Phi_n. \tag{1d}$$

***Remark 1.*** *In real-world applications, it is impractical to model the DHN based on equations (1a)-(1d). First, there are usually no measurement units for the pipelines and nodes in the DHN, making the inner operational data unavailable. Second, equations (1a)-(1d) will expose detailed information about the DHN, bringing serious privacy concerns [27]. Moreover, they have many variables and equations, increasing the computational cost.*

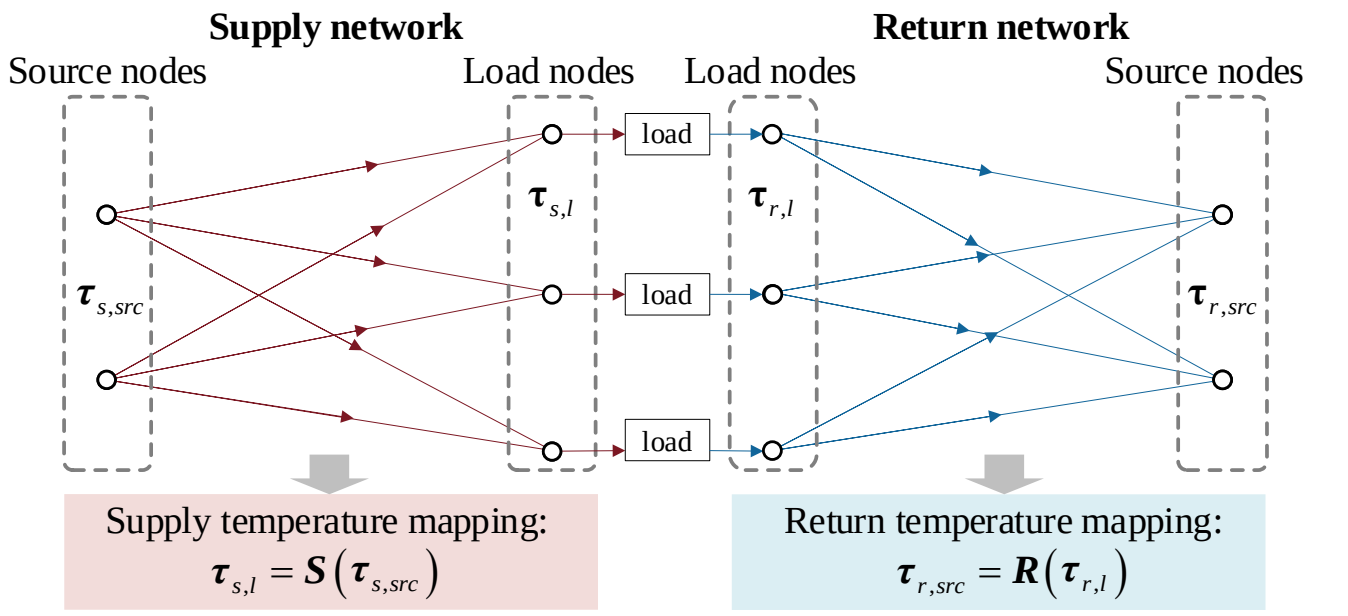

Fig. 2 Concept of the AGM of DHN.

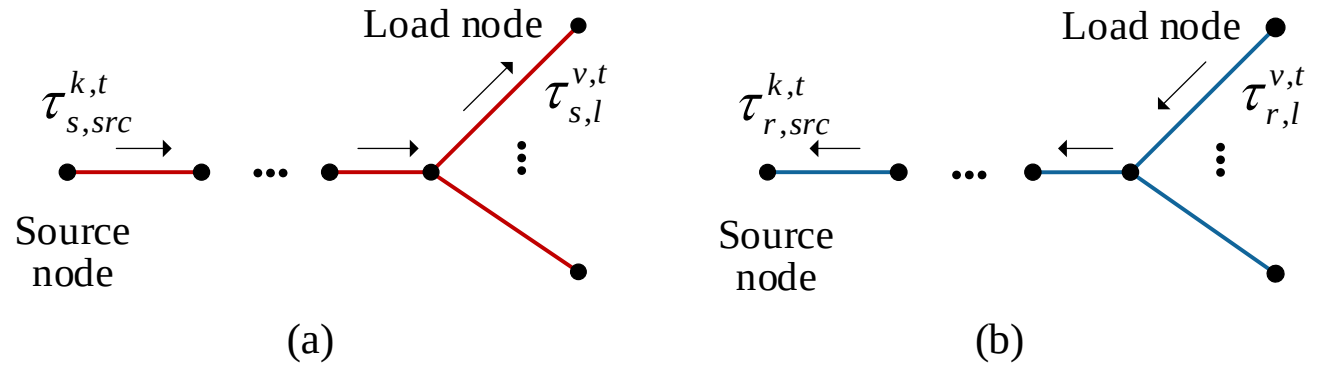

Fig. 3 Strcuture of single-source DHN: (a) Supply network; (b) Return network.

## B. Concept of the Aggregate Model

The concept of the AGM is shown in Fig. 2, which includes the supply network and the return network. In the supply network, the supply temperature of load nodes can be directly represented by the supply temperature of the source nodes. In the return network, the return temperature of the source nodes can be directly represented by the return temperature of the load nodes. The supply and return networks are coupled through the input power at the source node and the output power at the load node. The model of the heat load does not affect the derivation of the AGM since it is not included in the AGM. The proposed AGM consists of the STM and RTM, as

STM: $$\boldsymbol{\tau}_{s,l}^{t} = S\left(\boldsymbol{\tau}_{s,src}^{t}\right), \tag{2a}$$

RTM: $$\boldsymbol{\tau}_{r,src}^{t} = R\left(\boldsymbol{\tau}_{r,l}^{t}\right), \tag{2b}$$

wherein $S$ and $R$ are vector-valued functions, $S: \mathbb{R}^{(\Gamma+1)\times N_s} \to \mathbb{R}^{1\times N_l}$, $R: \mathbb{R}^{(\Gamma+1)\times N_l} \to \mathbb{R}^{1\times N_s}$ ; $\boldsymbol{\tau}_{s,src}^{t} \in \mathbb{R}^{(\Gamma+1)\times N_s}$ , $\boldsymbol{\tau}_{r,l}^{t} \in \mathbb{R}^{(\Gamma+1)\times N_l}$, $\boldsymbol{\tau}_{s,l}^{t} \in \mathbb{R}$, and $\boldsymbol{\tau}_{r,src}^{t} \in \mathbb{R}$, which are defined as

$$\boldsymbol{\tau}_{s,src}^{t} = \begin{bmatrix} \tau_{s,src}^{1,t-\Gamma} & \cdots & \tau_{s,src}^{N_s,t-\Gamma} \\ \vdots & \ddots & \vdots \\ \tau_{s,src}^{1,t} & \cdots & \tau_{s,src}^{N_s,t} \end{bmatrix}, \boldsymbol{\tau}_{r,l}^{t} = \begin{bmatrix} \tau_{r,l}^{1,t-\Gamma} & \cdots & \tau_{r,l}^{N_l,t-\Gamma} \\ \vdots & \ddots & \vdots \\ \tau_{r,l}^{1,t} & \cdots & \tau_{r,l}^{N_l,t} \end{bmatrix}, \tag{2c}$$

$$\boldsymbol{\tau}_{s,l}^{t} = \left[\tau_{s,l}^{1,t} \cdots \tau_{s,l}^{N_l,t}\right]^{\mathrm{T}}, \boldsymbol{\tau}_{r,src}^{t} = \left[\tau_{r,src}^{1,t} \cdots \tau_{r,src}^{N_s,t}\right]^{\mathrm{T}}. \tag{2d}$$

In practical engineering, we can use the measurements of the temperature at the source and load nodes to estimate the function $S(\cdot)$ and $R(\cdot)$. The detailed derivation of the STM and RTM is introduced in the following.

## C. Formation of AGM

First, we derive the AGM for single-source DHN, which is first introduced in [24]. Fig. 3 (a) gives a basic structure in a single-source radial DHN. Based on the recursion of the equation (1a), the supply temperature of the load node $v$ can be calculated as

$$\tau_{s,l}^{v,t} = \sum_{i=0}^{N_p^{k,v}} \left( a_s^{k,v,i} \tau_{s,src}^{k,t-\gamma_{agg}^{k,v}-i} \right) + b_s^{k,v} \tau_{amb} \quad \forall v \in \Phi_{ln}, \tag{3a}$$

wherein the aggregate parameter $a_s^{k,v,i}$, $b_s^{k,v}$, and $\gamma_{agg}^{k,v}$ are calculated as

$$a_s^{k,v,i} = \begin{cases} a_s^{v-1,v,0} a_s^{k,v-1,0}, & i = 0 \\ \sum_{n=0}^{1} a_s^{v-1,v,n} a_s^{k,v-1,i-n}, & 1 \le i \le v-k-1, \\ a_s^{v-1,v,1} a_s^{k,v-1,v-k-1}, & i = v-k \end{cases} \tag{3b}$$

$$b_s^{k,v} = 1 - \prod_{j\in\Phi_p^{k,v}} \left(1-\eta_p^j\right), \tag{3c}$$

$$\gamma_{agg}^{k,v} = \sum_{j\in\Phi_p^{k,v}} \gamma_p^j, \tag{3d}$$

wherein the parameters $a_s^{j-1,j,0}$ and $a_s^{j-1,j,1}$ are defined as

$$\begin{cases} a_s^{j-1,j,0} = \left(1-\eta_p^j\right)\alpha_p^j \\ a_s^{j-1,j,1} = \left(1-\eta_p^j\right)\left(1-\alpha_p^j\right) \end{cases} \quad j \in \Phi_p^{k,v}. \tag{3e}$$

The detailed derivation of (3a)-(3e) is given in [28].

Since the DHN under the constant flow and variable temperature control strategy is a linear time-invariant system (taking $\tau_{s,src}^{k,t}$ and $\tau_{amb}$ as the input and $\tau_{s,l}^{v,t}$ as the output), the superposition theorem holds for it. Therefore, the return temperature of the source node $k$ can be easily calculated as

$$\tau_{r,src}^{k,t} = \sum_{v\in\Phi_{ln}} \xi_r^{k,v} \sum_{i=0}^{N_p^{k,v}} \left( a_r^{k,v,i} \tau_{r,l}^{v,t-\gamma_{agg}^{k,v}-i} + b_r^{k,v} \tau_{amb} \right), \tag{3f}$$

wherein the parameter $a_r^{k,v,i}$, $b_r^{k,v}$, and $\xi_r^{k,v}$ are defined as

$$a_r^{k,v,i} = a_s^{k,v,i}, b_r^{k,v} = b_s^{k,v}, \xi_r^{k,v} = \frac{m_l^{k,v}}{\sum_{i\in\Phi_{ln}} m_l^{k,i}}. \tag{3g}$$

Based on the above, the AGM for the multi-source DHN can be easily derived based on the superposition theorem, as

$$\tau_{s,l}^{v,t} = \sum_{k\in\Phi_{sn}} \xi_s^{k,v} \left( \sum_{i=0}^{N_p^{k,v}} a_s^{k,v,i} \tau_{s,src}^{k,t-\gamma_{agg}^{k,v}-i} + b_s^{k,v} \tau_{amb} \right) \quad \forall v \in \Phi_{ln}, \tag{4a}$$

$$\tau_{r,src}^{k,t} = \sum_{v\in\Phi_{ln}} \xi_r^{k,v} \left( \sum_{i=0}^{N_p^{k,v}} a_r^{k,v,i} \tau_{r,l}^{v,t-\gamma_{agg}^{k,v}-i} + b_r^{k,v} \tau_{amb} \right) \quad \forall k \in \Phi_{sn}. \tag{4b}$$

In (4a), the coefficient $\xi_s^{k,v}$ is an constant representing the proportion of the water mass from the source node $k$ (i.e., $m_l^{k,v}$) to that flow into the load node $v$ (i.e., $\sum_{k\in\Phi_{sn}} m_l^{k,v}$); and the coefficient $\xi_r^{k,v}$ is an constant representing the proportion of the water mass from the load node $v$ (i.e., $m_{src}^{k,v}$) to that flow into the source node $k$ (i.e., $\sum_{v\in\Phi_{ln}} m_{src}^{k,v}$). The expressions of them are as

$$\xi_s^{k,v} = \frac{m_l^{k,v}}{\sum_{k\in\Phi_{sn}} m_l^{k,v}}, \xi_r^{k,v} = \frac{m_{src}^{k,v}}{\sum_{k\in\Phi_{ln}} m_{src}^{k,v}}. \tag{4c}$$

Because the water mass from multiple sources will mix before flowing into the load nodes, the proportions of the water mass from the source $k$ in $m_l^v$ are agnostic, leaving $\xi_s^{k,v}$ and $\xi_r^{k,v}$ without explicit expressions.

Equations (4a)-(4b) give the expressions of the STM and RTM defined in (2a) and (2b), which can be recast as

$$\tau_{s,l}^{v,t} = S^v\left(\boldsymbol{\tau}_{s,src}^t\right) = 1^{\mathrm{T}}\left(\tilde{\boldsymbol{a}}_s^v \circ \boldsymbol{\tau}_{s,src}^t\right)1 + \tilde{b}_s^v \tau_{amb} \ \forall v \in \Phi_{ln}, \quad (5a)$$

$$\tau_{r,src}^{k,t} = R^k\left(\boldsymbol{\tau}_{s,src}^t\right) = 1^{\mathrm{T}}\left(\tilde{\boldsymbol{a}}_r^k \circ \boldsymbol{\tau}_{r,l}^t\right)1 + \tilde{b}_r^k \tau_{amb} \ \forall k \in \Phi_{sn}, \quad (5b)$$

wherein ° is the Hadamard product (i.e., the element-wise product); $S^v(\cdot)$ and $R^k(\cdot)$ are the $v$th and $k$th element in $S(\cdot)$ and $R(\cdot)$, respectively; and the parameters $\tilde{\boldsymbol{a}}_s^k \in \mathbb{R}^{(\Gamma+1)\times N_s}$, $\tilde{\boldsymbol{a}}_r^v \in \mathbb{R}^{(\Gamma+1)\times N_l}$, $\tilde{b}_s^v \in \mathbb{R}$, and $\tilde{b}_r^k \in \mathbb{R}$ are defined as

$$\tilde{\boldsymbol{a}}_s^v = \begin{bmatrix} \tilde{a}_s^{1,v,\Gamma} & \cdots & \tilde{a}_s^{N_s,v,\Gamma} \\ \vdots & \ddots & \vdots \\ \tilde{a}_s^{1,v,0} & \cdots & \tilde{a}_s^{N_s,v,0} \end{bmatrix}, \tilde{\boldsymbol{a}}_r^k = \begin{bmatrix} \tilde{a}_r^{k,1,\Gamma} & \cdots & \tilde{a}_r^{k,1,\Gamma} \\ \vdots & \ddots & \vdots \\ \tilde{a}_r^{k,1,0} & \cdots & \tilde{a}_r^{k,N_l,0} \end{bmatrix}, \quad (5c)$$

$$\tilde{b}_s^v = \sum_{k\in\Phi_{sn}} \xi_s^{k,v} b_s^{k,v}, \tilde{b}_r^k = \sum_{v\in\Phi_{ln}} \xi_r^{k,v} b_r^{k,v}, \quad (5d)$$

wherein $\Gamma$, $\tilde{a}_s^{k,v,i}$, and $\tilde{a}_r^{k,v,i}$ are defined as

$$\Gamma = \max_{k\in\Phi_{sn}, v\in\Phi_{ln}} \left(\gamma_{agg}^{k,v} + N_p^{k,v}\right), \quad (5e)$$

$$\begin{aligned} \tilde{a}_s^{k,v,i} &= \begin{cases} \xi_s^{k,v} a_s^{k,v,i-\gamma_{agg}^{k,v}} & 0 \le i - \gamma_{agg}^{k,v} \le N_p^{k,v} \\ 0 & others \end{cases} \\ \tilde{a}_r^{k,v,i} &= \begin{cases} \xi_r^{k,v} a_r^{k,v,i-\gamma_{agg}^{k,v}} & 0 \le i - \gamma_{agg}^{k,v} \le N_p^{k,v} \\ 0 & others \end{cases} \end{aligned}. \quad (5f)$$

The definition in (5f) can be directly derived from (4a)-(4b). Taking (4a) as an example, for the temperature of the source node $k$, the coefficients of $\tau_{s,src}^{k,t-\gamma_{agg}^{k,v}-i}$, i.e., $a_s^{k,v,i}$, is nonzero only when $i = 0, \cdots, N_p^{k,v}$.

For the AGM (5a)-(5b), once we get the measurements $\boldsymbol{\tau}_{s,src}^t$, $\boldsymbol{\tau}_{r,l}^t$, $\tau_{s,l}^{v,t}$, $\tau_{r,src}^{k,t}$, and $\tau_{amb}$, the model parameters including $\tilde{\boldsymbol{a}}_s^v$, $\tilde{\boldsymbol{a}}_r^k$, $\tilde{b}_s^v$, and $\tilde{b}_r^k$ can be estimated. Taking the AGM of load node $v$ in the supply network as an example, based on the basic linear algebra knowledge, it can be deduced that when the measurement data has no error, the minimum measurement information required to estimate the model parameters is as

$$\tau_{s,l}^{v,t} \quad t = t_{end}, t_{end}-1, \ldots, t_{end} - (1+\Gamma)\cdot N_s \quad (6a)$$

$$\tau_{s,src}^{k,t} \quad \forall k \in \Phi_{sn}, t = t_{end}, \ldots, t_{end} - \Gamma - (1+\Gamma)\cdot N_s \quad (6b)$$

In (6a) and (6b), $t_{end}$ indicates the time period in which the last set of data resides; $N_s$ is the number of source nodes, which is usually less than 2; the value of $\Gamma$ depends on the scale of the heating network. For most DHN in the real-world, the value of $\Gamma$ is lower than 100. Therefore, the number of the required information is usually less than 303 sets in the ideal case. When the quality of the measurement data is poor, we need to increase the number of equations appropriately.

In addition, the time step has an impact on the accuracy of the node method model. A detailed analysis can be found in the paper [25]. The AGM is derived from the node method model, and hence the smaller the time step taken, the more accurate the AGM will be. In engineering practice, the time step of the measurement data of the DHN is often from 5 minutes to 30 minutes. Our test results show that the aggregate model based on 30-minute time step data has high precision and can meet the needs of practical problems such as economic dispatch.

***Remark 2.*** *(1) Although the STM and RTM are linear systems, as shown in (5a) and (5b), the inverse problem, i.e., the parameter estimation, is much more difficult because of the unknown structure of* $\tilde{\boldsymbol{a}}_s^v$ *and* $\tilde{\boldsymbol{a}}_r^k$*.The reason is that the unknown delay parameters of the DHN (*$\gamma_{agg}^{k,v}$*) and the number of pipelines between sources and loads (*$N_p^{k,v}$*) introduce the parameters to be estimated into the time index, as shown in (5f). Essentially, this is caused by the incomplete measurements of the DHN, i.e., the mass flow rates of pipelines (and the corresponding pipeline lengths and network topology) are not included in the measurements in engineering.*

*(2) In practice, we can usually only obtain temperature measurement data from the source and load nodes, resulting in many fundamental parameters being unidentifiable, such as the heat loss coefficient of pipelines (*$\lambda_p^j$*). The parameters in AGM are essentially a combination of fundamental parameters. Although the parameter estimation based on the AGM cannot derive the fundamental parameters, the AGM can describe the source-load relationship accurately, which is sufficient for practical applications such as economic dispatch and energy flow analysis.*

## III. Physics-Informed Robust Parameter Estimation of AGM

In this section, we develop the robust parameter estimation model for the AGM. Then, the delay parameter enumeration-based IRLS algorithm is proposed to solve this model efficiently.

### *A. Physics-Informed Robust Parameter Estimator*

*1) Robust estimation model*

The models defined in (5a)-(5f) show that $\tau_{s,l}^{v,t}$ is an affine function of $\tau_{s,src}^{k,t-\Gamma}$, $\tau_{s,src}^{k,t-\Gamma+1}$, ... , $\tau_{s,src}^{k,t}$, $k \in \Phi_{sn}$, and $\tau_{r,src}^{k,t}$ is an affine function of $\tau_{r,l}^{v,t-\Gamma}$, $\tau_{r,l}^{v,t-\Gamma+1}$, ... , $\tau_{r,l}^{v,t}$, $v \in \Phi_{ln}$. Therefore, we can model $\tau_{s,l}^{v,t}$ as a $M$-horizon linear regression model of $\tau_{s,src}^{k,t}$, $k \in \Phi_{sn}$ and $\tau_{r,src}^{k,t}$ as a $M$-horizon linear regression model of $\tau_{r,l}^{v,t}$, $v \in \Phi_{ln}$, respectively, wherein $M$ is defined as

$$M = \Gamma - \gamma_{agg}^{k,v}. \quad (7)$$

Based on this, the least squares estimator (LSE) can be used to obtain the parameters of the AGM, as

$$\forall v \in \Phi_{ln}: \ \min_{\tilde{a}_s^v, \tilde{b}_s^v} \ J_s^v\left(\tilde{\boldsymbol{a}}_s^v, \tilde{b}_s^v\right) = \frac{1}{2}\left(\boldsymbol{r}_s^v\right)^{\mathrm{T}} \boldsymbol{r}_s^v, \quad (8a)$$

$$\forall k \in \Phi_{sn}: \ \min_{\tilde{a}_r^k, \tilde{b}_r^k} \ J_r^k\left(\tilde{\boldsymbol{a}}_r^k, \tilde{b}_r^k\right) = \frac{1}{2}\left(\boldsymbol{r}_r^k\right)^{\mathrm{T}} \boldsymbol{r}_r^k. \quad (8b)$$

wherein the residual vectors $\boldsymbol{r}_s^v$ and $\boldsymbol{r}_r^k$ are defined as

$$\begin{aligned} \boldsymbol{r}_s^v &= \left[r_s^{v,1}, \cdots, r_s^{v,T}\right]^{\mathrm{T}} \\ &= \left[\tau_{s,l}^{v,1} - S^v\left(\boldsymbol{\tau}_{s,src}^1\right), \cdots, \tau_{s,l}^{v,T} - S^v\left(\boldsymbol{\tau}_{s,src}^T\right)\right]^{\mathrm{T}}, \end{aligned} \quad (8c)$$

$$\begin{aligned} \boldsymbol{r}_r^k &= \left[r_r^{k,1}, \cdots, r_r^{k,T}\right]^{\mathrm{T}} \\ &= \left[\tau_{r,src}^{k,1} - R^k\left(\boldsymbol{\tau}_{r,l}^1\right), \cdots, \tau_{r,src}^{k,T} - R^k\left(\boldsymbol{\tau}_{r,l}^T\right)\right]^{\mathrm{T}}. \end{aligned} \quad (8d)$$

In practical engineering, the quality of the measurements of

**Algorithm 1. The IRLS method for the HME.**

```
Initialize: set κ=1.345, Flag=1, i = 0, and tolerance ϵ.
Calculate an initial estimate (â_s^v, b̂_s^v)_0 by LSE, and calculate the
residual (r_s^v)_i by (8c) and scale (σ̂_s^v)_i by (9c).
While Flag
    Update i = i + 1.
    For t = 1 : T
        If |(r_s^{v,t})_{i-1}/(σ̂_s^v)_{i-1}| ≤ κ, set ω_i^t = 1;
        Else, set ω_i^t = κ/|(r_s^{v,t})_{i-1}/(σ̂_s^v)_{i-1}|.
    End
    Solve (â_s^v, b̂_s^v)_i = argmin_{(ã_s^v, b̃_s^v)_i ∈ Λ_s^v} Σ_{t∈T} ½ ω_i^t (r_s^{v,t}/(σ̂_s^v)_i)^2.
    Calculate the residual (r_s^v)_i by (8c) and scale (σ̂_s^v)_i by (9c).
    If ||(r_s^v)_i − (r_s^v)_{i-1}|| ≤ ϵ
        Set Flag=0.
    End
End While
```

**Algorithm 2. Delay parameter enumeration based IRLS method for the estimation of AGM.**

```
S0: Prescribe Δ^{k,v}, k ∈ Φ_sn, v ∈ Φ_ln.
S1. Estimation of STM:
For v = 1 : |Φ_ln|
    Generate all the combinations of the elements in Δ^{k,v}, k ∈
    Φ_sn, denoted as the set Ξ_1, Ξ_2 ⋯.
    For i = 1 : ∏_{k∈Φ_sn} |Δ^{k,v}|
        Prescribe the structure of a_s^v based on Ξ_i.
        Using the IRLS method to solve the estimator for the STM,
        (12), and denote the residual as J_s^v(Ξ_i).
    End
    Set Ξ_opt = argmin_{Ξ_i} J_s^v(Ξ_i).
End
S2. Estimation of RTM:
For k = 1 : |Φ_sn|
    Prescribe the structure of a_r^k based on Ξ_opt.
    Using IRLS method to solve the estimator for the RTM.
End
```

DHN cannot be guaranteed owing to many factors, such as environmental noise and poor communication quality. Hence, we use the Huber M-estimator (HME) to improve the robustness against the anomalies and outliers in the measurements. In the following, we only focus on (8a) for conciseness, but the method also applies to (8b).

Based on the HME, the LSE in (8a) is replaced by

$$\forall v \in \Phi_{ln}: \quad \min_{\tilde{\boldsymbol{a}}_s^v, \tilde{b}_s^v} \quad J_s^v\left(\tilde{\boldsymbol{a}}_s^v, \tilde{b}_s^v\right) = \sum_{t \in \mathrm{T}} J_s^{v,t}\left(\tilde{\boldsymbol{a}}_s^v, \tilde{b}_s^v\right), \tag{9a}$$

wherein $J_s^{v,t}(\boldsymbol{\alpha}_s^v, b_s^v)$ is the Huber loss, defined as

$$J_s^{v,t}\left(\tilde{\boldsymbol{a}}_s^v, \tilde{b}_s^v\right) = \begin{cases} \frac{1}{2}\left(r_s^{v,t}/\hat{\sigma}_s^v\right)^2 & \left|r_s^{v,t}/\hat{\sigma}_s^v\right| \le \kappa \\ \kappa\left|r_s^{v,t}/\hat{\sigma}_s^v\right| - \frac{1}{2}\kappa^2 & \left|r_s^{v,t}/\hat{\sigma}_s^v\right| > \kappa \end{cases}, \tag{9b}$$

wherein $\hat{\sigma}_s^v$ is the scale estimate of residuals, and $\kappa$ is a tuning constant (1.345 in this paper [29, 30]).

The scale estimate $\hat{\sigma}_s^v$ in (9b) is calculated by the median absolute deviation of the residuals, as

$$\hat{\sigma}_s^v = k \cdot \mathrm{median}\left(\left|\boldsymbol{r}_s^v - \mathrm{median}\left(\boldsymbol{r}_s^v\right)\right|\right), \tag{9c}$$

wherein $k = 1.4826$ for the normally distributed data. Please refer to paper [31] for the detailed derivation of $k$.

*2) Physics-informed estimator enhancement*

We notice that the physics below the AGM endows the parameters $\tilde{\boldsymbol{a}}_s^v$ and $\tilde{\boldsymbol{a}}_r^k$ with a special structure, including the normalization and sparsity.

First, as indicated by the expressions of $\tilde{a}_s^{k,v,i}$, $\tilde{a}_r^{k,v,i}$, $\tilde{b}_s^v$, $\tilde{b}_r^k$ in (5d), (5f) and the definitions (3b), (3c), (3g), (4c), we have the normalization constraints, as

$$1^{\mathrm{T}}\tilde{\boldsymbol{a}}_s^v 1 + \tilde{b}_s^v = 1 \quad \forall v \in \Phi_{sn}, \quad 1^{\mathrm{T}}\tilde{\boldsymbol{a}}_r^k 1 + \tilde{b}_r^k = 1 \quad \forall k \in \Phi_{ln}. \tag{10a}$$

Intuitively, the normalization constraints (10a) originate from the energy conservation law of the network. The detailed derivation of (10a) is given in [28].

Second, the equations in (5c) and (5f) indicate that the matrixes $\tilde{\boldsymbol{a}}_s^v$ and $\tilde{\boldsymbol{a}}_r^k$ should be sparse. These two features should be also embedded into the estimator in the form of constraints, as

$$\left(\tilde{\boldsymbol{a}}_s^v, b_s^v\right) \in \Lambda_s^v \quad \forall v \in \Phi_{sn}, \quad \left(\tilde{\boldsymbol{a}}_r^k, b_r^k\right) \in \Lambda_r^k \quad \forall k \in \Phi_{ln}, \tag{10b}$$

wherein $\Lambda_s^v$ and $\Lambda_r^k$ include the equations (10a) and the sparse structural properties.

Now, we can get the expression of the STM (5a) by solving (9a), (10a) and (10b). However, the equations (5e) and (7) indicate that the horizon $M$ will be very large if the scale of DHN is large, making the AGM tedious and not convenient for practical use. Fortunately, the equations (3b), (3e) and (3g) inspire a simplification method. Specifically, the equations (3b) indicate that the value of $a_s^{k,v,i}$ approximates 0 when $i$ approximates 0 or $k - v$ since $0 \le \alpha_p^j \le 1$ and $0 < 1 - \eta_p^j < 1$. Hence, we can use a truncated model with $M_{trc}$ horizons ($M_{trc} \ll M$) to approximate the original model. This can greatly reduce the computational complexity of parameter estimation and the required amount of measurement information. Our test results show that for most DHNs, the fourth-order model is sufficient to meet the accuracy requirements. Therefore, the model order can be adjusted appropriately based on actual needs. Adopting a truncated model is equivalent to using a new $\hat{\Lambda}_s^v$ (or $\hat{\Lambda}_r^k$) with a sparser structure for $\tilde{\boldsymbol{a}}_s^v$ (or $\tilde{\boldsymbol{a}}_r^k$) to approximate $\Lambda_s^v$ (or $\Lambda_r^k$). Hence, the constraints (10b) are replaced by

$$\left(\tilde{\boldsymbol{a}}_s^v, b_s^v\right) \in \hat{\Lambda}_s^v \quad \forall v \in \Phi_{sn}, \quad \left(\tilde{\boldsymbol{a}}_r^k, b_r^k\right) \in \hat{\Lambda}_r^k \quad \forall k \in \Phi_{ln}. \tag{11}$$

Finally, the HME for the STM of the AGM is formulated as

$$\forall v \in \Phi_{ln}: \quad \min_{\left(\tilde{\boldsymbol{a}}_s^v, \tilde{b}_s^v\right) \in \hat{\Lambda}_s^v} \quad J_s^v\left(\tilde{\boldsymbol{a}}_s^v, \tilde{b}_s^v\right) = \sum_{t \in \mathrm{T}} J_s^{v,t}\left(\tilde{\boldsymbol{a}}_s^v, \tilde{b}_s^v\right). \tag{12}$$

***Remark 3.*** *Since the estimator (12) integrates the physical properties of the DHN, including the normalization and sparsity constraints, its performance, including accuracy and robustness, will be enhanced. Theoretically, this is because the inherent structure of the AGM originating from the physical properties of DHN is enforced by the normalization constraints sparsity constraints and thus will not be destroyed by low-quality measurements. Numerical results will also verify this.*

## B. Physics-Enhanced Solution Algorithm

Numerous studies have shown that the classical IRLS algorithm is efficient in solving the HME [32, 33], as given in **Algorithm 1**. Usually, the global optimal solution can be obtained in a limited number of iterations. However, the IRLS method does not apply to (12) because of the presence of the structural constraint (11). As explained in ***Remark 2***, this constraint results in the locations and numbers of the non-zeros in $\tilde{\boldsymbol{a}}_s^v$ are unknown. Such a structural feature cannot be described by a closed-form expression, making the models (12) a nonlinear optimization problem with non-closed constraints.

Note that although the value of $\Gamma$ cannot be known in advance, it can be set to a large value bigger than the largest transmission delay of the DHN. This will make no difference except introducing more zeros in $\tilde{\boldsymbol{a}}_s^v$ and $\tilde{\boldsymbol{a}}_r^k$. Therefore, the fundamental difficulty is to keep the sparsity of $\tilde{\boldsymbol{a}}_s^v$ and $\tilde{\boldsymbol{a}}_r^k$. Specifically, the nonzeros in the $k$th column of $\tilde{\boldsymbol{a}}_s^v$ should be adjacent, and their number should be equal to $M_{trc}$.

***Remark 4.*** *Generally, we can introduce binary variables to provide explicit expressions for such structural constraints. However, the model (12) will turn into a large-scale mixed-integer nonlinear programming problem since the number of binary variables required for the structural description is large, making the model a computation-intensive problem, especially for large-scale networks. More seriously, the IRLS method will not be applicable because the convergence cannot be ensured after introducing binary variables.*

Fortunately, the definition in (5f) indicates that for the load node $v$, after the horizon number $M_{trc}$ is set, the nonzeros in the $k$ th column of $\tilde{\boldsymbol{a}}_s^v$ can be determined as $\tilde{a}_s^{k,v,\delta^{k,v}+M_{trc}}$,…, $\tilde{a}_s^{k,v,\delta^{k,v}+1}$, $\tilde{a}_s^{k,v,\delta^{k,v}}$ if the delay parameter $\delta^{k,v}$ is prescribed. In this situation, the STM estimator (12) and the corresponding RTM estimator turn to be a traditional HME. Inspired by this, we propose the delay parameter enumeration-based IRLS method for (12), as shown in **Algorithm 2**. The core of this method is to solve the estimator (12) under a set of prescribed delay parameters $\delta^{k,v}$, $k \in \Phi_{sn}$, $k \in \Phi_{ln}$ and then choose the one that has the smallest residuals as the estimated delay parameter. Assuming the set of the delay parameter $\delta^{k,v}$ to be selected is $\Delta^{k,v} = \{\delta_1^{k,v}, \delta_2^{k,v}, \cdots\}$, the number of times the HME needs to be solved for the STM and RTM, $K_s$ and $K_r$, respectively, can be calculated as

$$K_s = |\Phi_{ln}| \prod_{k \in \Phi_{sn}} |\Delta^{k,v}|, \tag{13a}$$

$$K_r = |\Phi_{sn}| \prod_{v \in \Phi_{ln}} |\Delta^{k,v}|. \tag{13b}$$

Theoretically, to get the best estimation of $\delta^{k,v}$, we need to compare the residual sum of STM and RTM under each combination of $\delta^{k,v}$. However, we note that the value of $K_r$ will be huge even for a small scale of DHN, resulting combinational explosion problem, while that of $K_s$ is modest. For example, for a DHN with 2 source loads and 10 load nodes, assuming that the number of delay parameters to be selected is 5 for each pair $(k,v)$, $k \in \Phi_{sn}$, $v \in \Phi_{ln}$, then we have $K_s$=250, $K_r$=19531250. Usually, we have $K_s \ll K_r$. Therefore, we propose to solve the STM estimators to obtain the delay parameters $\delta^{k,v}$ firstly, and then transfer the values of $\delta^{k,v}$ to the RTM estimators. By this method, the times that the estimator needs to be solved are as

$$K_{sum} = K_s + |\Phi_{ln}|. \tag{14}$$

Obviously, the proposed method dramatically reduces the computation burden, especially for a large scale of DHN. For example, for a DHN with 2 source loads and 100 load nodes, if the number of delay parameters to be selected is 5 for each pair $(k,v)$, $k \in \Phi_{sn}$, $k \in \Phi_{ln}$, we have $K_s$=2500 and $K_{sum}$=2600, meaning we only need to solve the estimators 2600 times.

Note that the delay parameters $\delta^{k,v}$ under fixed mass flow are constant, indicating we can use some methods to estimate their ranges a priori. Especially when the scale of the DHN is large or the time step used is small, the delay parameters are often large, and hence a more accurate delay time range needs to be determined in advance. A direct way to determine the delay time range is to calculate using the pipeline length and the water flow rate. However, the pipeline length and corresponding water flow rate are usually unknown. Another practical method is to compare and analyze the historical temperature curves to estimate the approximate delay time range. On the other hand, if the historical results are unavailable, we can relax the sparse constraints of the matrixes $\tilde{\boldsymbol{a}}_s^v$ and $\tilde{\boldsymbol{a}}_r^k$ to get a coarse estimation of the AGM, and then observe the nonzero parameters to determine the range of the delay parameters. Besides, in real-world engineering, the number of the sources in a DHN is usually small, about 1~4, far less than the number of load nodes. Therefore, the proposed solution strategy can efficiently avoid the combinational explosion problem and is practical for real-world applications.

## IV. Numerical Test

Two DHNs of different scales are simulated to verify the proposed method's effectiveness. Case I is based on a 7-node DHN to verify the accuracy of the AGM. Case II uses a real-world 42-node DHN located in Beijing, China. The operation data of Case I are obtained using the node method-based simulation. The operation data of Case II is from real-world measurements. The simulation platform is a laptop computer with an Intel i7 CPU and 16GB RAM. The programming environment is Matlab R2022a and Yalmip. Gurobi 10.0.1 is used to solve the (mixed integer) quadratic programming problem.

Two indexes are used to describe the data quality, including the standard deviation and the proportion of outliers. The proportion of outliers is defined as the proportion of data that deviates significantly from the general level in the total data. Three metrics were used to assess the goodness-of-fit of the model, including the root mean square error (RMSE), the mean absolute percentage error (MAPE), and the coefficient of determination ($R^2$), as follows:

$$\text{RMSE} = \sqrt{\frac{1}{1+T}\sum_{t=0}^{T}(\tilde{\tau}_t - \tau_t)^2},$$

$$\text{MAPE} = \frac{1}{1+T}\sum_{t=0}^{T}\left|(\tilde{\tau}_t - \tau_t)/\tau_t\right|,$$

$$R^2 = 1 - \sum_{t=0}^{T}(\tilde{\tau}_t - \tau_t)^2 \Bigg/ \sum_{t=0}^{T}\left(\tau_t - \frac{1}{1+T}\sum_{t=0}^{T}\tau_t\right)^2,$$

wherein $T$ is the number of samples, $\tilde{\tau}_t$ is the temperature calculated by AGM, and $\tau_t$ is the operation data.

### A. Case I: An Illustrative Network

The DHN in this case is shown in Fig. 4. The training and test data include 400 and 100 samples, respectively, with a

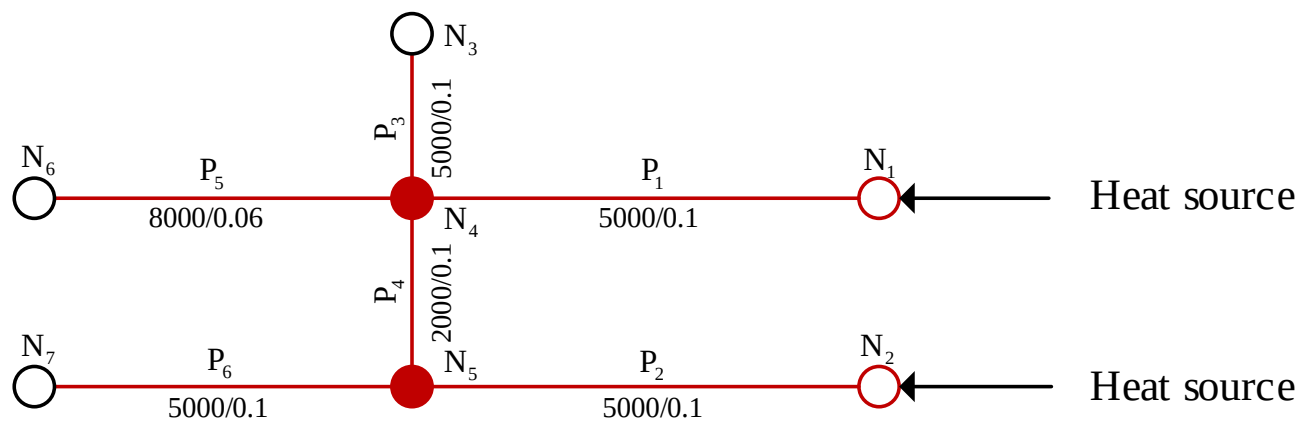

Fig. 4 The DHN structure of case 1.

TABLE I
SETTINGS OF TESTS

| Test No. | 1 | 2 | 3 | 4 | 5[a] | 6[b] |
|---|---|---|---|---|---|---|
| Standard deviation $\sigma$ | 0 | 1.0% | 1.0% | 1.0% | 1.0% | 1.0% |
| Proportion of outliers | 0 | 0 | 10% | 20% | 10% | 10% |

Note: [a] relaxing normalization constraints; [b] relaxing sparsity constraints.

resolution of 60 minutes. We perform several tests to verify the proposed method, the settings of which are given in Table I. Considering that the measurement error of the real-world DHN is usually below 1.0%, we add the Gaussian noises with 1.0% standard deviations to the exact data in Tests 2-6. In Tests 3-6, we add different proportions of outliers to data to investigate the robustness of the estimator. As mentioned in Section II, the theoretical values of the parameters in the AGM can be derived easily by (3a)-(4c). The model order in Tests 1-5 is set to 4 based on the theoretical results. Due to the page limitation, only the test results of the load node $N_6$ and the source node $N_2$ are analyzed in the following.

*1) Accuracy of the AGM*

The theoretical and estimated values of the parameters of the STM of $N_6$ and the RTM of $N_2$ are given in Table II and Table III, respectively. Test 1 indicates that the estimated parameters are in complete agreement with the theoretical values under noise-free data, verifying the theoretical accuracy of AGM. Although for the RTM of $N_2$, the value of $\gamma_{agg}^{2,7}$ in Test 1 is not correctly estimated, as shown in Table III, it does not affect the accuracy of the model parameters. The reason is that the theoretical values of the parameters $\tau_{r,l}^{7,t-1-i}$ are nonzeros only for $i$=0,1,2, so the parameters can be always precisely estimated when $\gamma_{agg}^{2,7}$ takes any value of 0,1, or 2. This also explains why the model parameters can be estimated accurately in other tests, although the delay parameters are not correctly estimated. Also, in Test 1, the values of $\tau_{s,src}^{1,t-2-i}$ ($i$=3) and $\tau_{r,l}^{7,t-1-i}$ ($i$=-1) are not strictly equal to the theoretical value 0. This comes from the numerical calculation-induced errors, whose impact is marginal and can be ignored.

Another interesting phenomenon is that for every node, the dominant values of the parameters $\tilde{a}_s^{k,v,i}$ (or $\tilde{a}_r^{k,v,i}$) are distributed in two adjacent positions, as shaded in blue in Table II and Table III. This proves the sparsity of the parameters $\tilde{\boldsymbol{a}}_s^v$ and $\tilde{\boldsymbol{a}}_r^k$, verifying the rationality of the proposed sparsity-based enhancement strategy.

*2) Robustness analysis*

In Tests 3-6, the outliers of varying proportions are added to the node temperature. We use the salt-and-pepper noise to simulate outliers [34-36], a common form of pulse noise. The probability distribution of the added noise is as

$$p(z)=\begin{cases} P_a & z=a \\ P_b & z=b \\ 1-P_a-P_b & z=1 \end{cases},$$

wherein, $z$ is the ratio of the noise data added to the original data; $P_a$ and $P_b$ are the probability of outliers. In the simulations, $a$ is set to 3 and $b$ is set to 0.3; and we set $P_a = P_b$. The proportion of outliers $P$ equals to $P_a + P_b$.

TABLE II
AGGREGATE PARAMETERS OF $N_6$ SOLVED BY HME IN CASE 1

| **Test No.** | $\gamma_{agg}^{1,6}$ | **Coefficients of $\tau_{s,src}^{1,t-2-i}$ (i.e., $\tilde{a}_s^{1,6,i}$)** | | | | | |
|---|---|---|---|---|---|---|---|
| | | ***i*=-2** | ***i*=-1** | ***i*=0** | ***i*=1** | ***i*=2** | ***i*=3** |
| Theoretical | 2 | /[a] | / | **0.38** | **0.23** | 0.026 | 0 |
| 1 | 2 | / | / | **0.38** | **0.23** | 0.026 | 0*[b] |
| 2 | 1 | / | 3.1e-4 | **0.38** | **0.22** | 0.023 | / |
| 3 | 1 | / | 3.7e-3 | **0.38** | **0.22** | 0.025 | / |
| 4 | 0 | 6.5e-3 | 3.7e-3 | **0.38** | **0.22** | / | / |
| 5 | 1 | / | 6.7e-3 | **0.38** | **0.22** | 0.022 | / |

| **Test No.** | $\gamma_{agg}^{2,6}$ | **Coefficients of $\tau_{s,src}^{2,t-2-i}$ (i.e., $\tilde{a}_s^{2,6,i}$)** | | | | | | $b_s^k$ |
|---|---|---|---|---|---|---|---|---|
| | | ***i*=-1** | ***i*=0** | ***i*=1** | ***i*=2** | ***i*=3** | ***i*=4** | |
| Theoretical | 2 | / | 1.4e-3 | **0.075** | **0.11** | 0.015 | / | 0.17 |
| 1 | 2 | / | 1.4e-3 | **0.075** | **0.11** | 0.015 | / | 0.17 |
| 2 | 2 | / | 6.9e-3 | **0.079** | **0.10** | 0.016 | / | 0.17 |
| 3 | 1 | 6.2e-3 | 8.2e-3 | **0.080** | **0.11** | / | / | 0.17 |
| 4 | 2 | / | 5.7e-3 | **0.079** | **0.11** | 0.019 | / | 0.17 |
| 5 | 1 | 3.4e-3 | 5.3e-3 | **0.078** | **0.11** | / | / | 0.094 |

| **Test No.** | $\gamma_{agg}^{1,6}$ | **Coefficients of $\tau_{s,src}^{1,t-1-i}$ (i.e., $\tilde{a}_s^{1,6,i}$)** | | | | | |
|---|---|---|---|---|---|---|---|
| | | ***i*=0** | ***i*=1** | ***i*=2** | ***i*=3** | ***i*=4** | ***i*=5** |
| 6 | 1 | 6.7e-4 | **0.38** | **0.22** | 0.023 | 5.1e-4 | 0* |

| **Test No.** | $\gamma_{agg}^{2,6}$ | **Coefficients of $\tau_{s,src}^{2,t-2-i}$ (i.e., $\tilde{a}_s^{2,6,i}$)** | | | | | | $b_s^k$ |
|---|---|---|---|---|---|---|---|---|
| | | ***i*=0** | ***i*=1** | ***i*=2** | ***i*=3** | ***i*=4** | ***i*=5** | |
| 6 | 2 | 5.5e-3 | **0.078** | **0.11** | 0.012 | 0* | 7.3e-3 | 0.17 |

[a] "/" denotes the model does not include this parameter, i.e., the default is 0;
[b] The parameter below 1e-4 is denoted as 0*.

The results in Table II and Table III indicate that the parameters $b_s^k$ and $b_r^k$ can be always precisely estimated in Tests 3-4. The estimation errors of $\tilde{a}_s^{k,v,i}$ are also small. However, the parameter $\tilde{a}_r^{2,3,i}$ is not accurately estimated. One reason is that the values of $\tilde{a}_r^{k,v,i}$ for some $v$ will be very small if the mass flow rate from the load node $v$ to the source node $k$ is small, i.e., the return temperature of this node $\tau_{r,l}^{v,t}$ contributes less to the that of the source node $\tau_{r,src}^{k,t}$. In this situation, the parameter $\tilde{a}_r^{k,v,i}$ is insensitive to the operation data and thus hard to be accurately estimated. Therefore, the results of Tests 3-4 indicate that the outliers could deteriorate the performance of the HME for insensitive parameters.

Fortunately, the error of the RTM caused by this situation is usually very small since the contribution of $\tau_{r,l}^{v,t}$ to $\tau_{r,src}^{k,t}$ is small. Specifically, for the return temperature of the source node $N_2$, the load nodes $N_3$, $N_6$, and $N_7$ and the ambient temperature contribute about 10.5%, 10%, 68.8%, and 11%, respectively. The goodness-of-fit metrics of Test 6, as given in Fig. 5, will also provide solid proof. Fig. 5 shows that the proposed HME has excellent goodness-of-fit metrics for all the nodes.

As shown in Fig. 5, the proposed method has significantly improved the estimation accuracy compared to the traditional

TABLE III
AGGREGATE PARAMETERS OF $N_2$ SOLVED BY HME IN CASE 1

| Test No. | $\gamma_{agg}^{2,3}$ | Coefficients of $\tau_{r,l}^{3,t-2-i}$ (i.e., $\tilde{a}_r^{2,3,i}$) $i$=-1 | $i$=0 | $i$=1 | $i$=2 | $i$=3 | $i$=4 |
|---|---|---|---|---|---|---|---|
| Theoretical | 2 | /[a] | 0*[b] | 2.5e-3 | **0.046** | **0.057** | / |
| 1 | 2 | / | 0* | 2.5e-3 | **0.046** | **0.057** | / |
| 2 | 2 | / | 0.012 | 0* | **0.065** | **0.054** | / |
| 3 | 1 | 7.6e-3 | 0.017 | 0* | **0.097** | / | / |
| 4 | 2 | / | 0.015 | 0* | **0.071** | **0.054** | / |
| 5 | 1 | 1.0e-3 | 0.013 | 0* | **0.090** | / | / |

| Test No. | $\gamma_{agg}^{2,6}$ | Coefficients of $\tau_{r,l}^{6,t-2-i}$ (i.e., $\tilde{a}_r^{2,6,i}$) $i$=-1 | $i$=0 | $i$=1 | $i$=2 | $i$=3 | $i$=4 |
|---|---|---|---|---|---|---|---|
| Theoretical | 2 | / | 6.8e-4 | **0.038** | **0.054** | 7.4e-3 | / |
| 1 | 2 | / | 6.8e-4 | **0.038** | **0.054** | 7.4e-3 | / |
| 2 | 2 | / | 0* | **0.038** | **0.051** | 0* | / |
| 3 | 1 | 0* | 0* | **0.040** | **0.051** | / | / |
| 4 | 2 | / | 0* | **0.033** | **0.048** | 0* | / |
| 5 | 1 | 0* | 0* | **0.040** | **0.046** | / | / |

| Test No. | $\gamma_{agg}^{2,7}$ | Coefficients of $\tau_{r,l}^{7,t-1-i}$ (i.e., $\tilde{a}_r^{2,7,i}$) $i$=-1 | $i$=0 | $i$=1 | $i$=2 | $i$=3 | $i$=4 | $b_r^k$ |
|---|---|---|---|---|---|---|---|---|
| Theoretical | 1 | / | 8.6e-3 | **0.47** | **0.21** | 0 | / | 0.11 |
| 1 | 0 | 0* | 8.6e-3 | **0.47** | **0.21** | / | / | 0.11 |
| 2 | 0 | 2.8e-3 | 9.9e-3 | **0.45** | **0.21** | / | / | 0.11 |
| 3 | 1 | / | 0.014 | **0.46** | **0.21** | 0* | / | 0.11 |
| 4 | 2 | / | / | **0.46** | **0.21** | 0* | 7.0e-3 | 0.11 |
| 5 | 1 | / | 8.2e-3 | **0.45** | **0.21** | 0* | / | 0.022 |

| Test No. | $\gamma_{agg}^{2,3}$ | Coefficients of $\tau_{r,l}^{3,t-2-i}$ (i.e., $\tilde{a}_r^{2,3,i}$) $i$=0 | $i$=1 | $i$=2 | $i$=3 | $i$=4 | $i$=5 |
|---|---|---|---|---|---|---|---|
| 6 | 2 | 5.4e-3 | 0* | **0.067** | **0.052** | 0* | 0* |

| Test No. | $\gamma_{agg}^{2,6}$ | Coefficients of $\tau_{r,l}^{6,t-2-i}$ (i.e., $\tilde{a}_r^{2,6,i}$) $i$=0 | $i$=1 | $i$=2 | $i$=3 | $i$=4 | $i$=5 |
|---|---|---|---|---|---|---|---|
| 6 | 2 | 0* | **0.041** | **0.048** | 1.7e-3 | 0* | 0* |

| Test No. | $\gamma_{agg}^{2,7}$ | Coefficients of $\tau_{r,l}^{7,t-1-i}$ (i.e., $\tilde{a}_r^{2,7,i}$) $i$=0 | $i$=1 | $i$=2 | $i$=3 | $i$=4 | $i$=5 | $b_r^k$ |
|---|---|---|---|---|---|---|---|---|
| 6 | 1 | 0.012 | **0.45** | **0.20** | 0* | 4.9e-3 | 1.9e-3 | 0.11 |

[a] "/" denotes the model does not include this parameter, i.e., the default is 0;
[b] The parameter below 1e-4 is denoted as 0*.

LSE. Especially, the values of $R^2$ of the LSE for the nodes $N_1$ and $N_2$ are negative, indicating that the model cannot capture data trends. In conclusion, the proposed HME for the AGM significantly improves the robustness to outliers.

*3) Effect of physics-informed structural constraints*

The effect of the physics-informed structural constraints can be revealed by comparing the results of Test 3 and Tests 5-6. The results of Test 5 indicate that the values of $b_s^k$ and $b_r^k$ cannot be accurately estimated after relaxing the normalization constraints, as shaded in grey in Table II and Table III.

The order of the model in Test 6 is set to 6 to make it more accurate, higher than Test 3. The results of Test 6 indicate that the model parameters can be estimated with reasonable accuracy after relaxing sparsity constraints, although the accuracy is sometimes lower than in Test 3. However, a potential risk of relaxing sparsity constraints is obtaining a physically unrealistic model, the most critical deficiency. For example, as shown in Table III, the nonzero locations of the parameters $\tilde{a}_r^{2,3,i}$ are not continuously distributed from $i$=-1 to 4, violating the heat transmission law of DHN. Besides, it should be noted that Test 6 also needs to estimate the range of the transmission delay of nodes before the parameter estimation to ensure the dimensions of the matrix $\tilde{a}_s^v$ and $\tilde{a}_r^k$ finite. Therefore, relaxing the sparsity constraints usually requires a higher-order model to ensure accuracy, which is another shortcoming. The above tests indicate that the effect of the physics-informed constraints is directly reflected in the numerical values of individual parameters. Although in some cases it does not show a huge improvement in accuracy, its role in the parameter stability and model interpretability cannot be ignored.

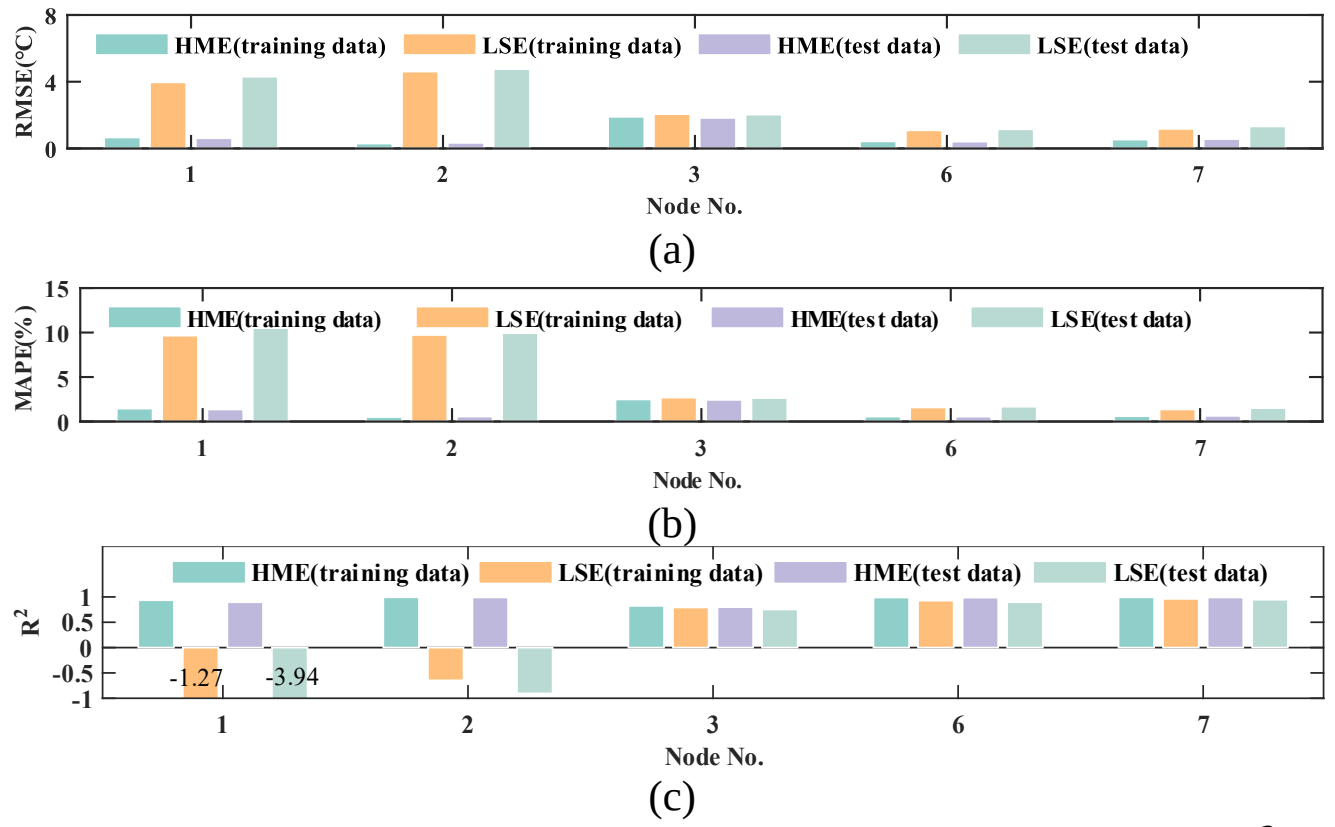

(a)

(b)

(c)

Fig. 5 The goodness-of-fit metrics in Test 4: (a) RMSE; (b) MAPE; (c) $R^2$.

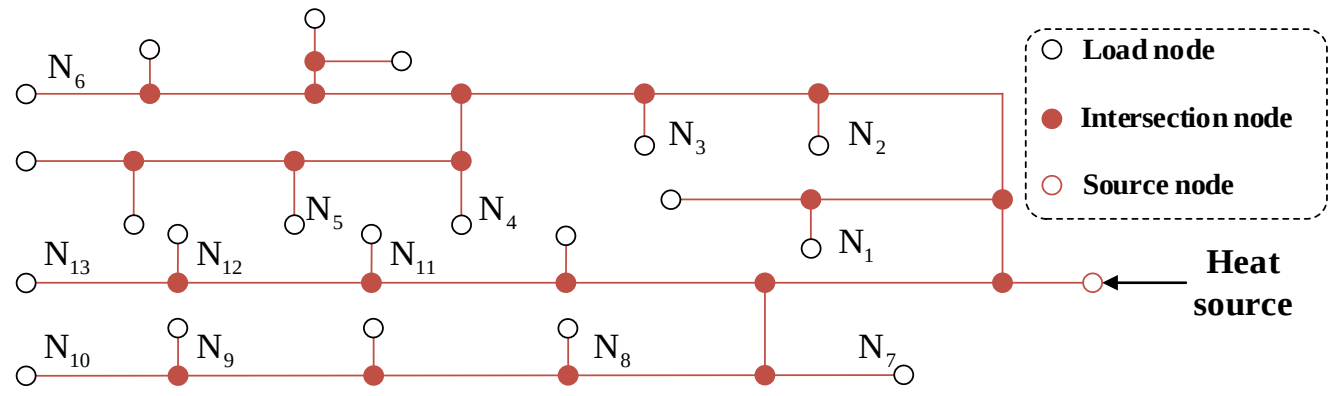

Fig. 6 Topology of DHN in case 2.

On the other hand, as stated in Section III, the results of Test 6 can provide a coarse estimation of the parameters, which can serve as the initial values for a more precise estimation of the delay parameters. This will efficiently improve the computational performance of the proposed model since the set of delay parameters is narrowed, especially when the historical information on the delay parameters is unavailable.

## *B. Case II: A Real-World Network*

The structure of DHN in Case II is given in Fig. 6. The training data and test data include 200 and 100 samples, respectively, with a resolution of 30 minutes. Due to incomplete real-world measurement data, only the AGMs of the partial supply network are tested, as labeled in Fig. 6. This is sufficient to prove the effectiveness of the AGM. To balance the computational cost and accuracy, we use the 3-horizon AGM.

The performance of the LSE and HME for the AGM are given in Fig. 7. The results indicate that both methods perform well while the HME is better than LSE for some nodes (e.g., $N_6$-$N_{11}$). Overall, the superiority of the HME cannot be fully exploited in this case because of the small proportions of outliers in the measurements (about 1% of the total data). Therefore, for the AGM, the gain of the HME compared to the LSE depends on the measurement quality.

Besides, the goodness-of-fit metrics of the training data of some nodes are worse than those of the test data, e.g., $N_3 - N_5$ and $N_{10}$-$N_{13}$. The reason is that there are more outliers in the training data of these nodes. Note that although the outliers have been filtered by HME, they will still have a negative impact on the goodness-of-fit metrics. Furthermore, Fig. 8 gives the results of the AGM on both the training and test data. In summary, we can conclude that the AGM can accurately describe the input-output characteristics of DHN, the accuracy of which is enough for the operation and control requirements of energy systems.

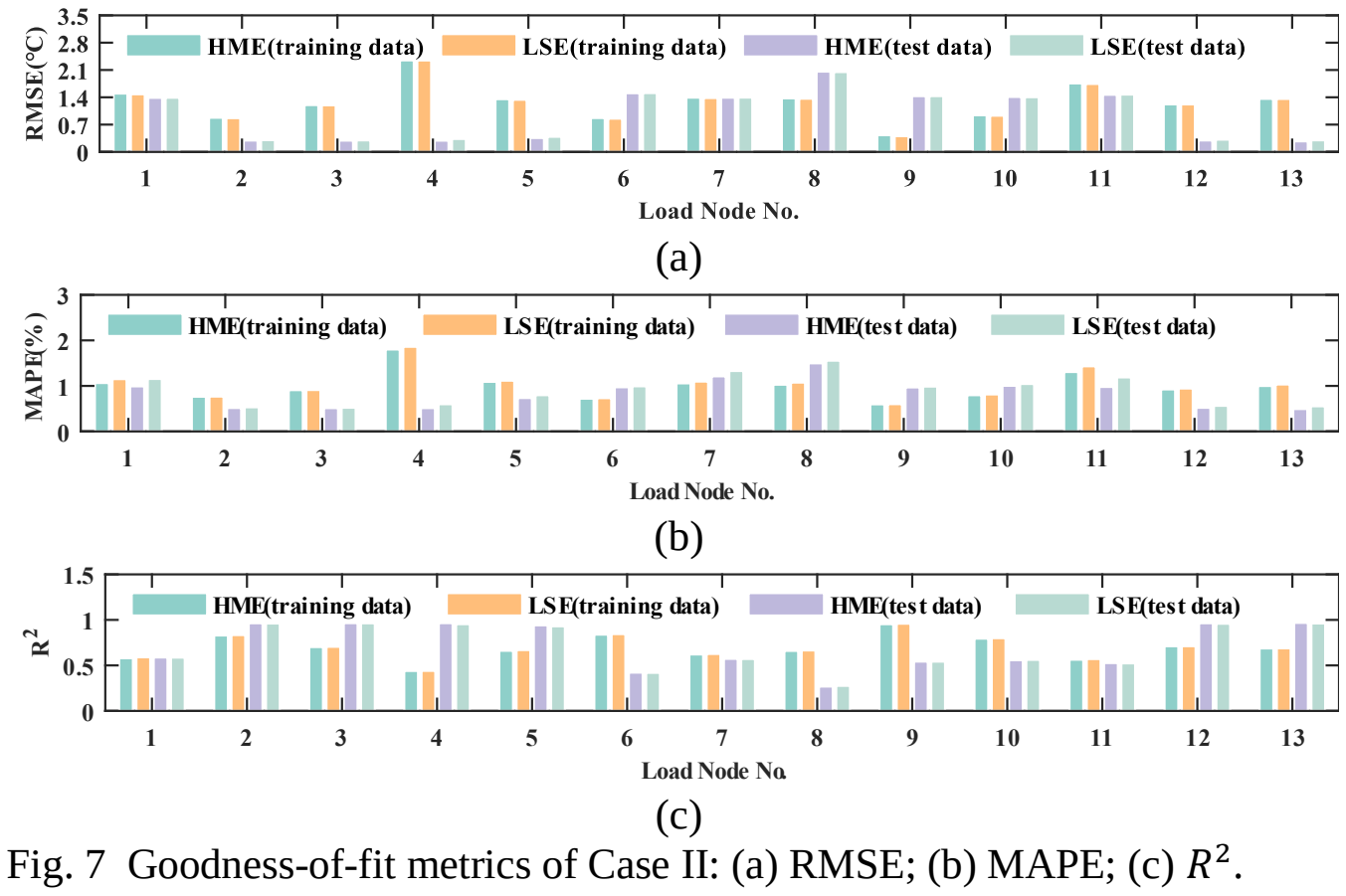

Fig. 7 Goodness-of-fit metrics of Case II: (a) RMSE; (b) MAPE; (c) $R^2$.

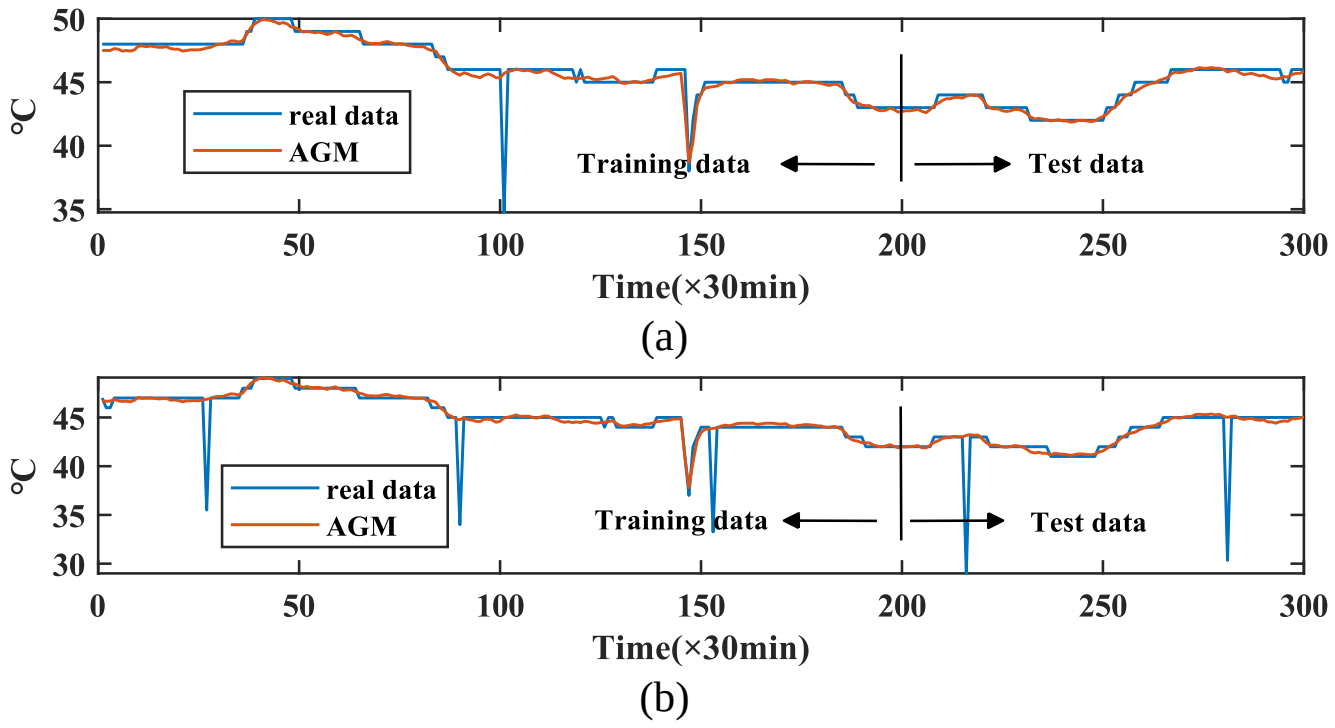

Fig. 8 Supply temperature of load nodes: (a) $N_2$; (b) $N_8$.

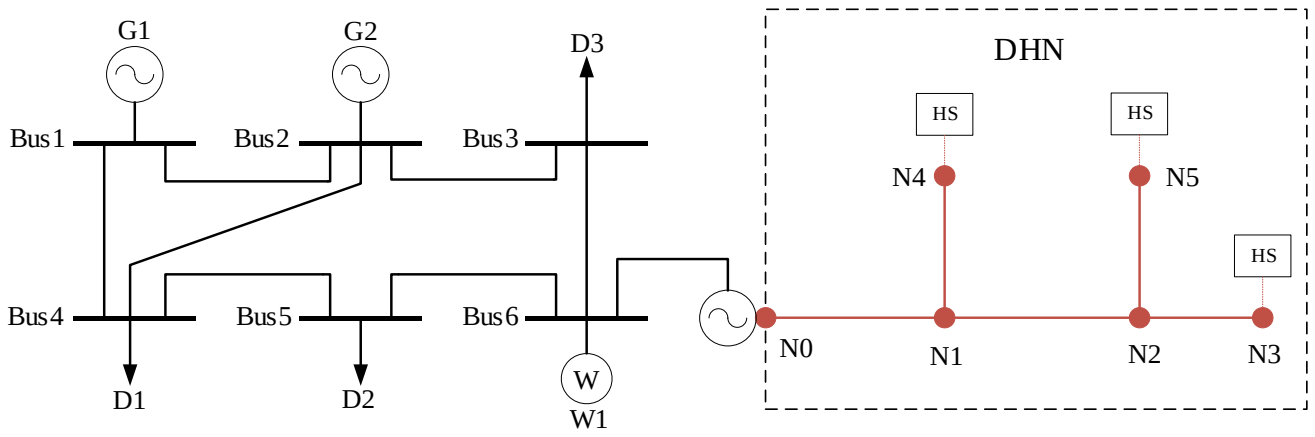

Fig. 9 Structure of IES in the small-scale system.

More detailed simulation results are provided in [28].

## *C. Verification of AGM in Economic Dispatch*

In this part, we investigate the performance of AGM in the economic dispatch problem of IES to verify its engineering application values. Two systems of different scales are investigated. For each system, two different subcases are simulated, respectively, in which Subcase I is based on the node method model; and Subcase II is based on the AGM. In the simulations, the dispatch period is set to 24 hours, the time interval is set to 1 hour, and the time resolution of the DHN and the building is set to 60 min. The initial state of the building is set to 21℃, and the upper and lower limits of the indoor temperature are set to 24℃ and 18℃, respectively.

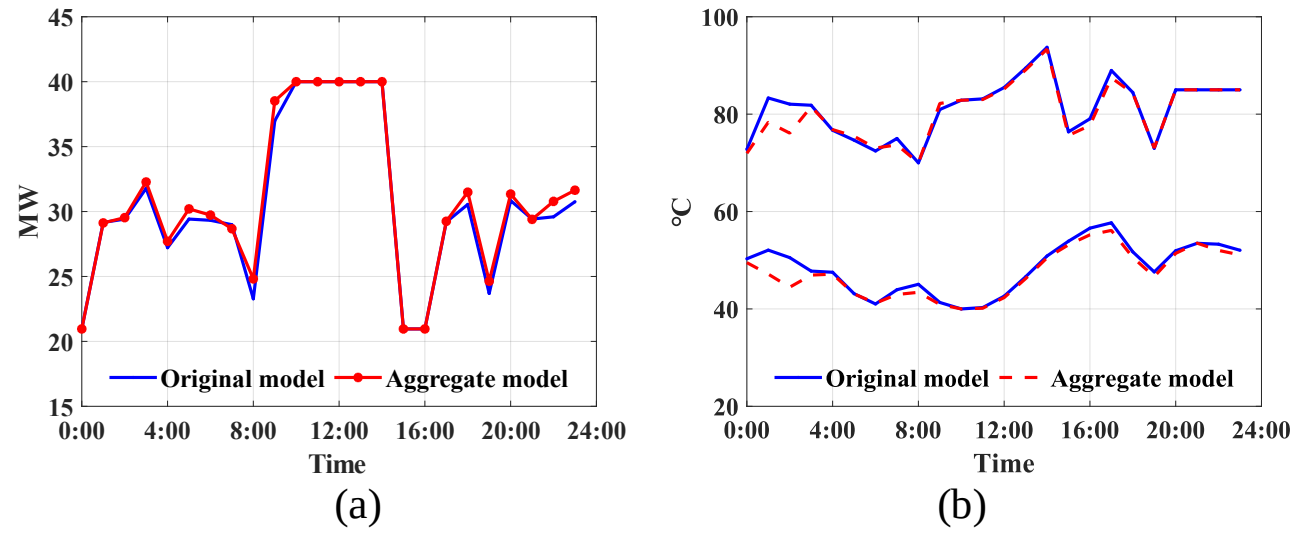

Fig. 10 (a) The heat power output of CHP in different models; (b) Supply/ return temperature of the source node in different models.

TABLE IV
RESULTS OF THE SMALL-SCALE SYSTEM

| DHN model | Cost (×1e3¥) | Solver time |
|---|---|---|
| Node method | 102.3912 | 0.1239s |
| AGM | 102.4989 | 0.0795s |
| Deviation | 1.05‰ | / |

TABLE V
RESULTS OF THE LARGE-SCALE SYSTEM

| DHN model | Cost (×1e6¥) | Heat energy of CHP (MWh) | Solver time |
|---|---|---|---|
| Node method [a] | \ | \ | ≥12h |
| AGM | 48.0103 | 70902 | 9190s |
| Node method [b] | 47.5433 | 68959 | 42.69s |
| AGM [b] | 47.6914 | 69853 | 13.93s |
| Deviation | 3.1‰ | 1.5% | / |

[a] The solver fails to finish within 12 hours;
[b] The binary variables denoting charging/discharging states are relaxed.

*1) Small-scale system*

The structure of the IES in this case is given in Fig. 9. The system consists of a 6-bus power system and a 6-node DHN. The dispatch model is a quadratic programming problem. The order of the AGM is set to 4. The dispatch results are shown in Fig. 10. In the two subcases, the heat power output of the CHP unit, namely the heat power injected into the DHN, is almost the same. Also, the heat energy output of the generation units in the two subcases is nearly the same. In Fig. 10 (b), the deviation between the supply (and return) temperatures of the two subcases is because this optimization problem has multiple solutions. In other words, as long as the difference between the supply temperature and the return temperature of the DHN remains consistent, the output heat power will remain consistent. Therefore, this optimization problem may have infinite solutions, the values of the objective function of which are the same. The dispatch costs and solver time are given in Table IV. The results indicate that the differences between the AGM and the node method model in the economic dispatch problem can be ignored.

*2) Large-scale system*

The IES in this case consists of a modified Polish 2383-bus power system and 20 DHNs. The power system contains 20 CHP units. Each CHP unit provides heat for a separate DHN. The topology of the DHN is modified from the DHN in [7], which contains 222 pipes and 223 nodes and provides heat for 112 heating exchange substations. The 20 DHNs have the same topology structure but different network parameters. Each DHN is equipped with a thermal storage tank at the source node. The dispatch model is a mixed-integer quadratic programming problem. The order of the AGM is set to 4. The simulation results, including the dispatch cost, output, and solver time, are given in Table V. The results show that the AGM model slightly increased the dispatch cost and the heat energy, but the amount is negligible. A potential reason is that the truncated AGM underestimates the supply/return temperature. It is worth mentioning that the solver time of the economic dispatch model based on the AGM is significantly reduced, revealing the significant computational advantages and the application potential of AGM, especially in large-scale systems.

In summary, the results demonstrate that the proposed AGM has good application prospects in the integrated energy optimization problem.

## V. Conclusion

The traditional DHN model relying on detailed pipeline modeling faces identifiability issues under limited measurements in engineering. To address this problem, this paper develops an AGM for the DHN that combines physical insights with data-driven techniques. The contributions of this work include the analytical expressions for the state relationship of source and load nodes, i.e., the AGM, a Huber M-estimator with physics-informed structural constraints for the parameter estimation of AGM, and an effective solution algorithm. Overall, our approach offers a promising solution for modeling the DHN, overcoming identifiability challenges and providing a foundation for integrated energy optimization.

Our future research will focus on 1) integrating buildings' thermal inertia into the AGM and 2) exploring the model identification of DHN under the variable flow control strategy.